\documentclass[11pt,fleqn]{article}

\usepackage{amsfonts}

\makeatletter
    \def\thebibliography#1{\section*{References\@mkboth
      {REFERENCES}{REFERENCES}}\list
      {[\arabic{enumi}]}{\settowidth\labelwidth{[#1]}\leftmargin\labelwidth
    \advance\leftmargin\labelsep
    \usecounter{enumi}}
    \def\newblock{\hskip .11em plus .33em minus .07em}
    \sloppy\clubpenalty4000\widowpenalty4000
    \sfcode`\.=1000\relax}
    \makeatother

\usepackage[utf8]{inputenc} 

\usepackage{geometry} 
\usepackage{graphicx} 
\usepackage{booktabs} 
\usepackage{array} 
\usepackage{paralist} 
\usepackage{verbatim} 
\usepackage{subfig} 

\usepackage{fancyhdr} 
\usepackage[nottoc,notlof,notlot]{tocbibind} 
\usepackage[titles,subfigure]{tocloft} 

\usepackage{amsmath} 
\usepackage{amssymb}
\usepackage{amsthm}
\newtheorem{theorem}{Theorem}[section]
\newtheorem{lemma}{Lemma}[section]

\newtheorem{corollary}{Corollary}[section]
\newtheorem{mydef}{Definition}[section]
\newtheorem{myrem}{Remark}[section]

\newtheorem{recall}{R}
\newtheorem*{mycav}{CAVEAT}

\usepackage{braket}
\usepackage{enumitem}

\usepackage{mathrsfs}

\title{{\bf NP} {\bf -} {\bf P} is not empty}
\author{Marius Constantin Ionescu\\
 North Vancouver, BC, Canada\\
\texttt{marius.c.ionescu@1wayfx.com}}

\date{} 

\begin{document}
\maketitle

\begin{abstract}
We present the {\it MEoP} problem that decides the existence of solutions to certain modular equations over prime numbers and show how this separates the complexity class {\bf NP} from its subclass {\bf P}.
\end{abstract}
 

\section{\normalfont \bf Introduction}
We concern with the existence of solutions to certain modular equations over prime numbers. We study the solution set of a specific instance of the following  generic functional equation with boundary conditions
\setlength{\belowdisplayskip}{5pt} \setlength{\belowdisplayshortskip}{5pt}
\setlength{\abovedisplayskip}{7pt} \setlength{\abovedisplayshortskip}{7pt}
\begin{equation}
\begin{cases}
F_1(x)^{(c^{x-2} \mod x)} \equiv F_2(x) \; \textrm{(mod $x$)} &  F:\mathbb{P} \rightarrow \mathbb{N}^2 , F(x) =(F_1(x),F_2(x)) \in \mathbb{Z}_x^2\\ 
x \in S, c \in \mathbb{N} & S \subset \mathbb{P}, c \textrm{ parameter}, F_1(x) \textrm{ generator of } \mathbb{Z}_x^*.
\end{cases}
\end{equation}

In particular, we study the equation instance $(F_1(x),F_2(x))=(\varphi_x,\pi_x), S = S_{(c,k)}$, where $\varphi_x$ depends on the number of generators of $ \mathbb{Z}_x^*$ (Definition 2.2), $\pi_x$ depends on the number of prime numbers less than or equal to $x$ (Definition 2.4), and $S_{(c,k)}$ is a set of prime numbers that is exponential with respect to the bit size of the pair $(c,k)$ (Definition 2.5).

 \indent Associated with that instance we introduce the {\it MEoP} problem that decides the non-emptiness of the solutions sets.  Specifically, {\it 'Given $(n,k)$, is it true that there is a field $\mathbb{Z}_p, p \in S_{(n,k)}$, in which the multiplicative inverse of n is the discrete logarithm of $\pi_p$ to the base $\varphi_p$'}. We show that the {\it MEoP} problem is in ${\bf NP}  {\bf -}  {\bf P}$, since there is no algorithm that solves the {\it MEoP} problem in less time than the cardinal numbers of the exponential sets $S_{(n,k)}$. Consequently, we have a separation of the complexity class {\bf NP} from its subclass {\bf P}.
\newline \indent The paper is organized as follows. In section 2, we fix the notation used and we state some lemmas concerning the solution set of the said equation. In section 3, we state the theorems and proofs of the main result.

\section{\normalfont \bf Notation and Preliminaries}

We let $\mathbb{P}$ denote the set of all odd prime numbers. $\mathbb{N}$ and $\mathbb{R}$ denote the set of natural numbers and real numbers respectively.  $\left\vert{M}\right\vert$ denotes the cardinal number of the set M. We let $\mathbb{S}$ denote the set of square numbers.
For $m,n \in \mathbb{N}$ and $x \in \mathbb{R}$, $\varphi(n) $ denotes the \textrm{\it totient} function, $\pi(x)$ denotes the \textrm{\it prime counting} function, ${\left\lfloor x \right\rfloor}$ denotes the {\it floor} function, and $gcd(m,n)$ denotes the {\it greatest common divisor} of $m$ and $n$. For $m,n \in \mathbb{N}$, $m \oplus n $ yields the integer whose base-2 representation is the bitwise XOR of the base-2 representation of the positive integers $m$ and $n$. 
If $p_1^{e_1}\cdots p_q^{e_q}$ is the {\it prime factorization} of $n \in \mathbb{N}$ then we let $P_n=\set{p_1^{e_1},\dots, p_q^{e_q}}$. The statements of a finite set $S$ of mathematical statements are {\it mutually exclusive} and {\it exhaustive} {\it iff} one and only one statement in $S$ is true.
\newline \indent  For $p \in \mathbb{P}$, $\mathbb{Z}_p$ denotes the field of integers modulo $p$ and $\mathbb{Z}_p^*$ denotes the multiplicative group of $\mathbb{Z}_p$. We let $G_p$ denote the set of generators of $\mathbb{Z}_p^* $. We let $Q_p$ denote the set of quadratic residues of $\mathbb{Z}_p^* $.  We use $\textrm{\it ind}_g \,a \pmod{p}$ to denote the discrete logarithm (index) of $a$ with respect to the base $g$ modulo $p$. We use $o_p(g)$ for the multiplicative order of $g \in \mathbb{Z}_p^* $. We let $\left(\frac{a}{p}\right)$ denote the Legendre symbol of $a$ modulo $p$.
\newline \newline
\noindent   Let $l>10^2$  be an arbitrary but fixed integer.
\newline \newline 
Recall that
\begin{recall} 
For $p \in \mathbb{P},p>l$, $ \left\vert{G_p}\right\vert =\varphi(p-1) < \frac{p}{2} $. 
\end{recall}
\begin{recall} 
If $p$ is a prime $ p >5$, then the generators of $\mathbb{Z}_p^* $ are not consecutive \normalfont \cite{mon}.
\end{recall}
\begin{recall} 
The gaps between consecutive generators of $\mathbb{Z}_p^* $ are at most $p^{\frac{1}{4}+o(1)}$ \normalfont \cite{bu1} \cite{bu2}.
\end{recall}
\begin{recall} 
For $p \in \mathbb{P}, p>l$ and $g \in \mathbb{Z}_p^* \setminus \set{1,p-1}$,
\begin{enumerate}[nolistsep, font=\normalfont,label=(\roman*)]
\item
$g \in G_p$ {\it iff} $\left(\frac{g}{p}\right)=-1$, $g^{\frac{p-1}{p_i}} \not \equiv 1 \pmod{p}$, for any odd prime divisor $p_i$ of $(p-1)$
\item
$g \in Q_p$ {\it iff} $\left(\frac{g}{p}\right)=1$. In particular, $g \in Q_p$ if $g \in \mathbb{S}$
\item
 $g \notin G_p \cup Q_p$ {\it iff} $o_p(g) \ne p-1$ and $o_p(g)$ is not a divisor of ${\frac{p-1}{2}}$.
\end{enumerate}
\end{recall}
\begin{recall} 
For $n \in \mathbb{N}$ and $p,q \in \mathbb{P}$, if $2<n<p,q$ then $n^{p-2} \mod p \ne n^{q-2} \mod q$.
\end{recall}

\noindent 
As consequences of R.4 we have:
\begin{lemma}
For $p \in \mathbb{P}, p>l$, and $e \in \mathbb{Z}_p^* $, if $e \in \mathbb{S}$ then in deciding whether $e \in Q_p$, no $\mathbb{Z}_p $ field operations are required to be performed directly on $e$. If $e \notin \mathbb{S}$ then in deciding whether $e \in G_p$ or $e \in Q_p$ or $e \notin G_p \cup Q_p$, $\mathbb{Z}_p $ field operations are required to be performed directly on $e$, given that $\left(\frac{e}{p}\right)$ or $o_p(e)$ must be computed.
\end{lemma}

\begin{mydef}
We let $I = \set{(n,k) | n,k \in \mathbb{N},n>2,k > max \set{n+1,l}^{max \set{n+1,l}}}$.
\end{mydef}

\noindent   For $p \in \mathbb{P}, p>l$, 

\begin{mydef}
We let $\varphi_p = min \set{g|g \in G_p, g \ge \varphi(p-1)}$.
\end{mydef}

\begin{mydef}
\setlength{\belowdisplayskip}{0pt} \setlength{\belowdisplayshortskip}{0pt}
\setlength{\abovedisplayskip}{0pt} \setlength{\abovedisplayshortskip}{0pt}
We let 
\begin{equation*}
e_p = \begin{cases}
p  \oplus (p-\pi(p)) & \textrm{\qquad if $p  \oplus (p-\pi(p)) < p$}\\ 
2^{\left\lfloor \log_2 p \right\rfloor} \oplus p  \oplus (p-\pi(p)) & \textrm{\qquad if $p  \oplus (p-\pi(p)) > p$}.
\end{cases}
\end{equation*}
\end{mydef}  

\begin{mydef}
\setlength{\belowdisplayskip}{0pt} \setlength{\belowdisplayshortskip}{0pt}
\setlength{\abovedisplayskip}{0pt} \setlength{\abovedisplayshortskip}{0pt}
We let 
\begin{equation*}
\pi_p = \begin{cases}
((e_p)^{p-2} \mod p) & \textrm{\qquad if $((e_p)^{p-2} \mod p) \ne \varphi_p$}\\ 
((e_p)^{p-2} \mod p) +1 & \textrm{\qquad if $((e_p)^{p-2} \mod p) = \varphi_p$}.
\end{cases}
\end{equation*}
\end{mydef}  

\begin{myrem}
\normalfont By R.1, R.2 and R.3, $\sqrt{p}<\varphi_p<p$ . Since $\pi(p) \notin \set{0,1,p-1}$, $e_p<p$  and $e_p \notin \set{0,1,p-1}$ and thus, $ \pi_p<p$  and $ \pi_p \notin \set{0,1,p-1}$.
\end{myrem}

\begin{lemma}
For all $p \in \mathbb{P}, p>l$, the decision whether $\pi_p \in \mathbb{S}$ requires $\mathbb{Z}_p$ field operations to be performed directly on $e_p$ and $\varphi(p-1)$. That is, deciding whether $\pi_p \in \mathbb{S}$ requires $\mathbb{Z}_p $ field operations.
\begin{proof}
For each $p \in \mathbb{P}, p>l$, $\pi_p$ must be evaluated numerically in order to decide whether $\pi_p \in \mathbb{S}$, given that the membership problem in the set $\mathbb{S}$ implies a given positive integer as input. By definition, $\pi_p$ computation implies the computation of the multiplicative inverse of $e_p$ in $\mathbb{Z}_p^* $  and the computation of the generator $\varphi_p$. The former computations require $\mathbb{Z}_p $ field operations, given that $(e_p)^2 \not \equiv 1 \pmod p$ and $\varphi(p-1) $ has to be checked as a generator of $\mathbb{Z}_p^* $ since $\varphi_p = min \set{g|g \in G_p, g \ge \varphi(p-1)}$. Therefore, $\mathbb{Z}_p $ field operations are required  in order to decide whether $\pi_p \in \mathbb{S}$. 
\end{proof}
\end{lemma}

\begin{myrem}
\normalfont For all $p \in \mathbb{P}, p>l$, deciding whether $\pi_p \in \mathbb{S}$ not only requires $\mathbb{Z}_p$ field operations, but also relies on the computation of the prime factorization of $p-1$ (e.g., to compute $\varphi(p-1)$ and to test an element in $\mathbb{Z}_p^* $ as a generator) and the computation of $p  \oplus (p-\pi(p))$. The former computations were omitted, since they were additional to the result.
\end{myrem}  

\noindent  For  $(n,k) \in I$, 

\begin{mydef}
We let  $S_{(n,k)} = \set{p | p \in \mathbb{P}, max \set{n+1,l} < p < k }$.
\end{mydef} 

\begin{mydef}
We let  $S_{(n,k)} ^{sol} = \set{p | p \in S_{(n,k)} , (\varphi_p)^{{(n^{p-2} \mod p)}} \equiv \pi_p \pmod{p} }$. That is, $S_{(n,k)} ^{sol}$ is the solution set of the equation $(\varphi_x)^{{(n^{x-2} \mod x)}} \equiv \pi_x \pmod{x}$ restricted to the set $S_{(n,k)}$.
\end{mydef}

\begin{mydef}
We let MEoP  denote the problem: 'Given $(n,k) \in I$, is it true that $\left\vert{S_{(n,k)} ^{sol}}\right\vert > 0$' (MEoP is the initialism for Modular Equations over Primes). Equivalently: 'Given $(n,k) \in I$, is it true that there is a field $\mathbb{Z}_p, p \in S_{(n,k)}$, in which the multiplicative inverse of n in $\mathbb{Z}_p^*$  is the discrete logarithm of $\pi_p$ to the base $\varphi_p$'.
\end{mydef}

\begin{myrem}
\normalfont In each $\mathbb{Z}_p^*$, the equation $(\varphi_p)^{x} \equiv \pi_p \pmod{p}$ has an unique solution, given that $\varphi_p \in G_p$. Hence,  given $(n,k) \in I$, it is impossible to justify the decision $p \notin S_{(n,k)}^{sol}$ on the non-existence of a solution to the said equation. 
\end{myrem}

\noindent It is easy to see that the following lemmas hold.

\begin{lemma}
For $p \in \mathbb{P}, p>l$, the following statements are mutually exclusive and exhaustive:
\begin{enumerate}[nolistsep, font=\normalfont,label=(\roman*)]
\item
 $\pi_p \in G_p$ iff $gcd(p-1,\textrm{\it ind}_{\varphi_p} \,\pi_p \pmod{p}) = 1$
\item
 $\pi_p \in Q_p$ iff $gcd(p-1,\textrm{\it ind}_{\varphi_p} \,\pi_p \pmod{p})=2k, k \in \mathbb{N}$
\item
 $\pi_p \notin G_p \cup Q_p$ iff $gcd(p-1,\textrm{\it ind}_{\varphi_p} \,\pi_p \pmod{p}) =2k+1, k \in \mathbb{N}$.
\end{enumerate}
\end{lemma}

\noindent Therefore, if $p \in S_{(n,k)}^{sol}$, i.e., $n^{p-2} \mod p = \textrm{\it ind}_{\varphi_p} \,\pi_p \pmod{p}$, we have:
 
\begin{lemma}
For $(n,k) \in I$, if $p \in S_{(n,k)}^{sol}$ then the following statements are mutually exclusive and exhaustive:
\begin{enumerate}[nolistsep, font=\normalfont,label=(\roman*)]
\item
 $\pi_p \in G_p$ iff $gcd(p-1,{n^{p-2} \mod p}) = 1$
\item
 $\pi_p \in Q_p$ iff $gcd(p-1,{n^{p-2} \mod p})=2k,k \in \mathbb{N}$
\item
 $\pi_p \notin G_p \cup Q_p$ iff $gcd(p-1,{n^{p-2} \mod p}) =2k+1, k \in \mathbb{N}$.
\end{enumerate}
\end{lemma}

\noindent As consequences of the previous lemmas, we have:
\begin{lemma}
For $(n,k) \in I$, and $p \in S_{(n,k)}$,  the following statements are mutually exclusive and exhaustive:
\begin{enumerate}[nolistsep, font=\normalfont,label=(\roman*)]
\item
 $p \in S_{(n,k)}^{sol}$ and $\pi_p \in G_p$ and $gcd(p-1,{n^{p-2} \mod p}) = 1$
\item
  $p \in S_{(n,k)}^{sol}$ and $\pi_p \in Q_p$ and $gcd(p-1,{n^{p-2} \mod p})=2m$, for $m \in \mathbb{N}$
\item
  $p \in S_{(n,k)}^{sol}$ and $\pi_p \notin G_p \cup Q_p$ and $gcd(p-1,{n^{p-2} \mod p}) =2m+1$, for $m \in \mathbb{N}$,
\item
  $p \notin S_{(n,k)}^{sol}$ and $\pi_p \in G_p$ and $gcd(p-1,{n^{p-2} \mod p}) = 1$
\item
  $p \notin S_{(n,k)}^{sol}$ and $\pi_p \in Q_p$ and $gcd(p-1,{n^{p-2} \mod p})=2m$, for $m \in \mathbb{N}$
\item
  $p \notin S_{(n,k)}^{sol}$ and $\pi_p \notin G_p \cup Q_p$ and $gcd(p-1,{n^{p-2} \mod p}) =2m+1$, for $m \in \mathbb{N}$
\item
   $p \notin S_{(n,k)}^{sol}$ and $\pi_p \in G_p$ and $gcd(p-1,{n^{p-2} \mod p}) \ne 1$
\item
   $p \notin S_{(n,k)}^{sol}$ and $\pi_p \notin G_p$ and $gcd(p-1,{n^{p-2} \mod p}) = 1$
\item
   $p \notin S_{(n,k)}^{sol}$ and $\pi_p \in Q_p$ and $gcd(p-1,{n^{p-2} \mod p}) \ne 1,2m$, for $m \in \mathbb{N}$
\item
   $p \notin S_{(n,k)}^{sol}$ and $\pi_p \notin Q_p$ and $gcd(p-1,{n^{p-2} \mod p})=2m$, for $m \in \mathbb{N}$
\item
   $p \notin S_{(n,k)}^{sol}$ and $\pi_p \notin G_p \cup Q_p$ and $gcd(p-1,{n^{p-2} \mod p}) \ne 1,2m+1$, for $m \in \mathbb{N}$
\item
   $p \notin S_{(n,k)}^{sol}$ and $\pi_p \in G_p \cup Q_p$ and $gcd(p-1,{n^{p-2} \mod p}) =2m+1$, for $m \in \mathbb{N}$.
\end{enumerate}
\end{lemma}

\section{\normalfont \bf Results}

\noindent Associated with each instance  $(n,k) \in I$ of the {\it MEoP} problem, we concern with the complexity of the decision whether $\left\vert{S_{(n,k)} ^{sol}}\right\vert > 0$ and show how this separates, via the {\it MEoP} problem, the complexity class {\bf NP} from its subclass {\bf P}.
\newline To achieve this result we state and prove the following results:
\begin{itemize}[nolistsep]
\item[-]  [Theorem 3.1] Given $(n,k) \in I$, for all $p \in S_{(n,k)}$, the decision whether $p \in S_{(n,k)}^{sol}$  requires $\mathbb{Z}_p$ field operations;
\item[-]  [Theorem 3.2] There is no determinstic algorithm that solves the {\it MEoP} problem in less than exponential time
\item[-]  [Theorem 3.3] The {\it MEoP} problem is not in ${\bf P}$;
\item[-]  [Theorem 3.4] The {\it MEoP} problem is in ${\bf NP}$;
\item[-]  [Theorem 3.5]   The MEoP problem is in ${\bf NP} {\bf -} {\bf P}$.
\end{itemize}

\noindent The following theorem shows that, with arbitrary given $n \in \mathbb{N}$, in the decision whether the equation $(\varphi_x)^{(n^{x-2} \mod x)} \equiv \pi_x \pmod{x}$  has a solution in $S_{(n,k)}$, there is no triplet $(n, p, \textrm{ the generator } \varphi_p)$, $p \in S_{(n,k)}$, that can infer a structural relationship among other $q \in S_{(n,k)}$ to allow shortcuts to checking each element in $S_{(n,k)}$, one by one. For each $p \in S_{(n,k)}$, whether $p \in S_{(n,k)} ^{sol}$ can not be inferred or computed outside $\mathbb{Z}_p$. The decision whether the equation $(\varphi_x)^{(n^{x-2} \mod x)} \equiv \pi_x \pmod{x}$  has a solution in $S_{(n,k)}$, is a posterior computation in each $\mathbb{Z}_p$.

On the one hand, in each $\mathbb{Z}_p$, the {\it 3}-tuple $(\varphi_p,\pi_q,\textrm{\it ind}_{\varphi_p} \,\pi_p \pmod{p})$ exists independently of any arbitrarily given $n \in \mathbb{N}$. On the other hand, with an arbitrary given $n \in \mathbb{N}$, in each $\mathbb{Z}_p$, $p>n$, the {\it 2}-tuple $(n,n^{p-2} \mod p)$  exists independently of the internal {\it 3}-tuple $(\varphi_p,\pi_q,\textrm{\it ind}_{\varphi_p} \,\pi_p \pmod{p})$. The decision whether the {\it 3}-tuple $(\varphi_p,\pi_q,\textrm{\it ind}_{\varphi_p} \,\pi_p \pmod{p})$ and the {\it 2}-tuple $(n,n^{p-2} \mod p)$ agree on the last component is a strictly internal decision in each field $\mathbb{Z}_p$. 

\begin{theorem}
For all pairs $(n,k) \in I$ and for all $p \in S_{(n,k)}$, the decision whether $p \in S_{(n,k)}^{sol}$  requires $\mathbb{Z}_p $ field operations.
\end{theorem}
\begin{proof}
Suppose, by way of contradiction, that there is a pair $(\bar{n},\bar{k}) \in I$ and there is a prime $\bar{p} \in S_{(\bar{n},\bar{k})}$ such that $\bar{p} \in S_{(\bar{n},\bar{k})}^{sol}$ is decidable without requiring $\mathbb{Z}_p $ field operations. By Lemma 2.5, for $\bar{p} \in S_{(\bar{n},\bar{k})}$, one and only one statement from (i) to (xii) must be true. Thus, one and only one of the statements $\pi_{\bar{p}} \in G_{\bar{p}}$, $\pi_{\bar{p}} \in Q_{\bar{p}}$, $\pi_{\bar{p}} \notin G_{\bar{p}} \cup Q_{\bar{p}}$ must be true, as a substatement of the said statement. By our supposition and by Lemma 2.1, the said true substatement must be $\pi_{\bar{p}} \in Q_{\bar{p}}$ with the property that $\pi_{\bar{p}} \in Q_{\bar{p}} \cap \mathbb{S}$ and hence, we have the decision $\pi_{\bar p} \in \mathbb{S}$ without requiring $\mathbb{Z}_p $ field operations. We note that the said statement must be either (ii) or (v) or (ix), depending on the decision whether $\bar{p} \in S_{(\bar{n}, \bar{k})}^{sol}$ and the value of $gcd(\bar p-1,{n^{\bar p-2} \mod \bar p})$, given that $\pi_{\bar p} \in \mathbb{S}$. 
Therefore, since the decision $\pi_{\bar p} \in \mathbb{S}$ did not require $\mathbb{Z}_p $ field operations, we get a contradiction to Lemma 2.2.
\end{proof}

\noindent   Let $(n,k)$  be  arbitrary in $I$.
\begin{corollary}
For all combinations of primes C in $S_{(n,k)}$, there is no single operation (test or property) to decide simultaneously the membership in $S_{(n,k)}^{sol}$ of all primes p in C without without requiring $\mathbb{Z}_p $ field operations for each prime p in C. That is, whether $p \in S_{(n,k)}^{sol}$ holds for a prime p has nothing to do with whether $p' \in S_{(n,k)}^{sol}$ holds for another prime $p'$ and there is no shortcut to the existential search of $S_{(n,k)}$.
\end{corollary}
\begin{proof}
Suppose, by way of contradiction, that there exists a combination of primes $C^*$ in $S_{(n,k)}$ such that there exists an operation that decides  simultaneously the membership in $S_{(n,k)}^{sol}$ of all primes p in $C^*$ without requiring $\mathbb{Z}_p$ field operations for each prime p in $C^*$. Now, for each prime $p$ in $C^*$  for which no $\mathbb{Z}_p$ field operations were required (by the said operation) to decide its membership in $S_{(n,k)}^{sol}$, we get a contradiction to Theorem 3.1, since deciding whether $p \in S_{(n,k)}^{sol}$ requires $\mathbb{Z}_p$ field operations.
\end{proof}

\noindent We let $\mathcal{A}_{MEoP}$ denote the set of all deterministic algorithms $\mathcal{A}$ that solve the {\it MEoP} problem.

\begin{theorem}

For all inputs $(n,k) \in I$ and for all algorithms $\mathcal{A} \in \mathcal{A}_{MEoP}$, if $\mathcal{A}$ outputs 'NO' on input $(n,k)$ then $\mathcal{A}$ requires $\mathbb{Z}_p $ field operations, for each $p \in S_{(n,k)}$. That is, $\mathcal{A}$ must check each $p \in S_{(n,k)}$, one by one. There are no shortcuts to exhaustive search of $S_{(n,k)}$.
\end{theorem}
\begin{proof}
Suppose, by way of contradiction, that there is an input $(\bar{n},\bar{k}) \in I$ and there is an algorithm $\mathcal{\bar A} \in \mathcal{A}_{MEoP}$ such that,  if $\mathcal{\bar A}$ outputs 'NO' on input $(\bar{n},\bar{k})$, $\mathcal{\bar A}$ does not require $\mathbb{Z}_p $ field operations, for each $p \in S_{(\bar{n},\bar{k})}$. Let $U_{(\bar{n},\bar{k})}$ be the set of all $\bar p \in S_{(\bar{n},\bar{k})}$ for which no $\mathbb{Z}_{\bar p} $ field operations were required by $\mathcal{\bar A}$ to output 'NO' on input $(\bar{n},\bar{k})$. By our supposition the set $U_{(\bar{n},\bar{k})}$ is not empty, for all $\bar p \in U_{(\bar{n},\bar{k})}$, $\bar p \notin S_{(\bar{n},\bar{k})}^{sol}$ (i.e., the set $S_{(\bar{n},\bar{k})}^{sol}$ is empty since $\mathcal{\bar A}$ outputs 'NO' on input $(\bar{n},\bar{k})$) and no $\mathbb{Z}_{\bar p} $ field operations were required in the decision $\bar p \notin S_{(\bar{n},\bar{k})}^{sol}$. Therefore, we get a contradiction to Theorem 3.1 which states that, for all ${\bar p} \in S_{(\bar{n},\bar{k})}$, the decision whether $\bar p \in S_{(\bar{n},\bar{k})}^{sol}$ requires $\mathbb{Z}_{\bar p} $ field operations.
\end{proof}

\noindent As consequences of the previous theorems, we have: 
\begin{enumerate}[nolistsep, font=\normalfont,label=(\roman*)]
\item  Whether $(\varphi_p)^{{(n^{p-2} \mod p)}} \equiv \pi_p \pmod{p} $  holds in a field $\mathbb{Z}_p$ has {nothing to do} with whether $(\varphi_q)^{{(n^{q-2} \mod q)}} \equiv \pi_q \pmod{q} $  holds in another field $\mathbb{Z}_q$
\item  There is {no collective behavior} (shortcuts) and there is {no yet-to-be-known property} (structural results) to decide whether the said equations hold in multiple fields $\mathbb{Z}_r$ {simultaneously}
\item  The existential search of $S_{(n,k)}$ is {mandatory} in the decision whether $\left\vert{S_{(\bar{n},\bar{k})} ^{sol}}\right\vert > 0$.
\end{enumerate}

\begin{mycav}
\normalfont In the context of the Theorem 3.1 and the Theorem 3.2, the statement '\textit {Individual checking has little to do with collective checking and thus, the results of this paper are invalid}' is erroneous.

Generically, let T1 denote any theorem equivalent to Theorem 3.1 and let T2 denote any theorem equivalent to Theorem 3.2. that relate to a decision problem EP. Clearly, if T1 holds than T2 holds and, contrapositively, if T2 does not hold then T1 does not hold (i.e., there exist elements in the search space of EP such that they can be decided in a different way from the way stated  in T1). 

Therefore, it is a logical error to apply the same logical reasoning to a problem EP by assuming that T1 holds and T2  does not hold. Any parallelism between {\it MEoP} problem and a problem EP for which  a shortcut exists (i.e., T2 does not hold) is erroneous. For instance, consider the following EP: '\textit {Decide whether a monic polynomial equation $p(x)=0, p(x) \in \mathbb{Z}[x]$ has integer solutions}'. By Rational Root Theorem, the membership problem in integer solution set is easy and hence, T2 and T1 do not hold (the exhuaustive search of all integers is not mandatory since only the divisors of the constant term are candidates). An erroneous approach would be:
\vspace{-3mm}
\begin{quote}
 {\normalfont Given $m \in \mathbb{Z}$ whether $p(m)=0$ may look difficult to decide without performing exponentiation upon $m$ (which is a part of $p$). Thus, one could assume the theorem T1: '\textit {There is no integer $m$ such that $p(m)=0$ is decidable without performing exponentiation upon $m$}'. However, the membership problem of integer solution set is easy and, as here, using exactly the same logical argument, one could argue that EP requires mandatory exhautive search.}
\end{quote}

\end{mycav}

\noindent The following theorem rules out the existence of any algorithm for deciding the {\it MEoP} problem in polynomial time. There are no shortcuts to exponential exhaustive search in solving the {\it MEoP}  problem.
\begin{theorem}
The MEoP problem is not in ${\bf P}$.
\end{theorem}
\begin{proof}
By Theorem 3.2, there is no algorithm that solves the {\it MEoP} problem in less time then the cardinal numbers of the sets $S_{(n,k)}$. Since each set $S_{(n,k)}$ is  exponential with respect to the bit size of the pair $(n,k)$, there is no algorithm that solves the {\it MEoP} problem in polynomial time. That is, the {\it MEoP} problem is not in ${\bf P}$.
\end{proof}
\begin{theorem}
The MEoP problem is in ${\bf NP}$.
\end{theorem}
\begin{proof}
If $p \in \mathbb{P},p>l$, let consider the certificate $C_p = \set{ P_{p-1}, \pi(p)}$. Observe that, if $(n,k) \in I$ and $p \in S_{(n,k)}$, $C_p$ has polynomial size with respect to the input instance $(n,k)$. Let $\mathcal{V}$ be an algorithm with the following properties:
\begin{enumerate}[nolistsep]
\item Input: $((n,k),C_p)$ 
\item Check $C_p$ using \cite{aks} and \cite{dus}
\item Compute $\varphi_p,\pi_p$ and $n^{p-2} \mod p$
\item Check $(\varphi_p)^{(n^{p-2} \mod p)} \equiv \pi_p \pmod{p}$ 
\item Output: 'solution set is not empty'  or  'non-emptiness of the solution set is not verified'.
\end{enumerate}
Since AKS primality test, $\varphi(p-1)$ computation (when the unique prime factorization of $p-1$ is given), a generator recognition in $\mathbb{Z}_p^* $  (when the unique prime factorization of $p-1$ is given) and modular inverse computation are polynomial, the algorithm $\mathcal{V}$ is polynomial. On input $((n,k),C_p)$ that satisfies $(\varphi_p)^{(n^{p-2} \mod p)} \equiv \pi_p \pmod{p}$, for $p \in S_{(n,k)}$,  $\mathcal{V}$ recognizes $(n,k)$ as 'solution set is not empty' and operates in polynomial time on input $((n,k),C_p)$. On input $((n,k),C_p)$ that does not satisfy $(\varphi_p)^{(n^{p-2} \mod p)} \equiv \pi_p \pmod{p}$, for $p \in S_{(n,k)}$, $\mathcal{V}$ recognizes $(n,k)$ as 'non-emptiness of the solution set is not verified'.
\end{proof}
\begin{theorem}
 The MEoP problem is in ${\bf NP} {\bf -} {\bf P}$.
\end{theorem}
\begin{proof}
By Theorem 3.3, the {\it MEoP} problem is not in ${\bf P}$, and by Theorem 3.4, the {\it MEoP} problem is in {\bf NP}. Since ${\bf P}$ is a subclass of the complexity class ${\bf NP}$, the {\it MEoP} problem is in ${\bf NP} {\bf -} {\bf P}$.
\end{proof}

\section{\normalfont \bf Final Remark}

\noindent The complexity class containment of the {\it MEoP} problem is directly related to the cardinality of the sets $S_{(n,k)}$. If we change the exponential aspect of the sets $S_{(n,k)}$ to  polynomial or infinite then the {\it MEoP} problem class containment changes to {\bf P} or undecidable (with a 1-way {\it yes} certificate) respectively.

\end{document}